%Paper: cond-mat/9402094
%From: Andrew P. Gould <gould@payne.mps.ohio-state.edu>
%Date: Tue, 22 Feb 94 07:37:22 EST

\input phyzzx
\hoffset=0.375in

\def\vecx{{\bf x}}
\def\vecy{{\bf y}}
\def\veck{{\bf k}}
\def\kx{\psi}
\def\ky{{\bf k \cdot y}\over 2}
\def\calf{{\cal F}}
\def\cpp{C_{p p}}
\def\var{\sigma^2}

\def\grk{{\xi}}
\def\eps{{\epsilon}}
\twelvepoint
\font\bigfont=cmr17
\centerline{\bigfont Weak Lensing By Nearby Structures}
\bigskip
\centerline{\bf Andrew Gould and Jens Villumsen}
\smallskip
\centerline{Department of Astronomy, Ohio State University, Columbus, OH 43210}
\smallskip
\centerline{E-mail gould@payne.mps.ohio-state.edu or jens@payne.mps.ohio-state}
\bigskip
\singlespace
\centerline{\bf Abstract}

	Weak gravitational lensing due to nearby structures,
such as the Coma cluster, and the Local Supercluster can be expected to
 polarize images
of distant galaxies by ${\cal O}(0.2\%\Omega)$ with coherence over scales of
tens of square degrees.   The Sloan Survey, which will image $\gsim 10^4$
galaxies $\rm deg^{-2}$ over $\pi$ steradians, should be sensitive to
polarizations of $\sim 0.1\% A^{-1/2}$, where $A$ is the area in square
degrees.  By measuring the polarization, one could determine $\Omega$
in local structures and compare this value to that derived from a
variety of other techniques.

Subject Headings: gravitational lensing -- large scale structure of the
universe

\endpage

\chapter{Introduction}
\normalspace

	Weak lensing, the distortion of images by a gravitational field
without the
creation of multiple images, is a potentially powerful tool for studying the
large-scale inhomogeneities of the universe (Kristian 1967; Gunn 1967).
In essence, when rays from a distant galaxy pass by an overdensity in
the matter distribution, the observed images will be elongated slightly
tangentially with respect to the center of the perturbation
(Lynds \& Petrosian 1989; Soucail et al.\ 1987;
Tyson, Valdes, \& Wenk 1990; Fort et al.\ 1991).  Independent perturbations
along the line of sight add stochastically (Blandford et al.\ 1991;
Miralda-Escud\'e 1991; Kaiser 1992).

	Early searches for weak lensing outside of rich
clusters (Kristian 1967; Valdes, Tyson \& Jarvis 1983) met with negative
results.  However, Mould et al.\ (1994) have reported a tentative
detection of a $2.8\%\pm 0.4\%$
polarization in a random high-latitude field based
on a very deep $10'$ square $r$ band CCD image.  The reported mean
polarization is well below the errors in the measurement of the
ellipticities of individual
galaxies on which the determination is based.

	Previous weak lensing studies have generally focused on distant
structures, either known clusters or field structures
at redshifts of several tenths.  There is a good reasons for this.
For fixed physical separation between a mass concentration and a given
line of sight and for a source at an infinite distance,
the strength of lensing scales as the distance to the lens.
For fixed
distance, the lensing declines with angular separation.
Hence, lensing is easiest to observe by imaging structures that just fit on
a single CCD image.   To image an entire cluster (diameter $\sim 3 h^{-1}$
Mpc) on a large CCD ($10'$), the cluster must be at $z\gsim 0.3$.

	However, a great deal could be learned if it were possible to measure
the weak lensing due to local structures, such as the Local Supercluster
or the Coma cluster.  Weak lensing is sensitive primarily to the total
mass in a given structure.  We have an enormous amount of information about
the mass distribution of local structures that is not available for more
distant structures.  For example, we know the peculiar motions of many galaxies
in the Local Supercluster, and even of galaxies and clusters of galaxies
at somewhat greater distances.  We have much more detailed information
about the gas distribution in the Coma cluster than we do of more distant
clusters.  Hence by measuring the weak lensing associated with these
structures we can both gain new insight into the structures that we understand
the best and also gain an external check for other methods of estimating
masses and mass distributions.

	In this {\it Letter} we show that the Sloan Survey (Gunn \& Knapp 1993)
is ideally suited to measure weak lensing induced by local structures in
the north galactic cap.  In \S\ 2, we show that the weak lensing polarization
can be measured with an accuracy $\sigma \sim (\Delta/7 N)^{1/2}$ where
$\Delta$ is the accuracy of the measurement of the ellipticities of individual
galaxies and $N$ is the number of galaxies measured.  In \S\ 3, we estimate
that for a 1 square degree patch, the Sloan Survey can be used to measure
polarizations to an accuracy $\sigma\sim 0.1\%$.  In \S\ 4, we show that
local structures such as the Virgo cluster produce weak lensing of
$\sim 0.2\%\Omega$ and that these structures are coherent over tens
 of square degrees.  The Coma cluster produces much stronger lensing.
In \S\ 5, we indicate that the Sloan Survey will also provide useful
information on lensing by more distant structures.  In \S\ 6, we discuss
the calibration of systematic effects.

\chapter{Detectability of Weak Lensing}
%\normalspace

	The distortion of images in the weak lensing limit can be parameterized
by a complex polarization $\grk$.  This polarization results in a displacement
in the complex ellipticity from the object, $\gamma$, to the complex
ellipticity of the image, $\eps$.  That is to lowest order
$$\grk = \eps - \gamma \eqn\grkdef$$
where
$$ \eps\equiv {I_{xx} - I_{yy} - 2 i I_{xy}\over I_{xx} + I_{yy}};
\qquad \gamma\equiv{O_{xx} - O_{yy} - 2 i O_{xy}\over O_{xx} + O_{yy}},
\eqn\grkdeftwo$$
and where
$I_{xx}$, $I_{yy}$, and $I_{xy}$ are the three second moments of the
image and $O_{xx}$, $O_{yy}$, and $O_{xy}$ are the three second moments of the
object.

	Consider a set of $N$ objects with complex ellipticities, $\gamma_j$.
First suppose that the $\gamma_j$ all have the same modulus $|\gamma|$, but
have random phases.  That is the $\gamma_j$ are random points on a circle
centered at the origin.  If all these images are subjected to the same
lensing polarization, then the resulting ellipticities of the images
$\eps_j$ will
be points on a circle which is displaced from the origin by $\grk$.
If the $\eps_j$ are measured with perfect accuracy, then real and imaginary
parts of $\grk$ can be determined with an accuracy
$\sigma=[\VEV{Re(\gamma)^2}/N]^{1/2}=|\gamma|/(2N)^{1/2}$.
If the real and imaginary parts of the $\eps_j$ can each be measured only to
an accuracy $\Delta/2^{1/2}$, then $\sigma$ increases to
$$\sigma = \biggl({|\gamma|^2 + \Delta^2\over 2 N}\biggr)^{1/2}.\eqn\sigone$$
Now suppose that the moduli of the ellipticities are distributed as
$f(|\gamma|)$ between 0 and 1, with $\int d|\gamma|\,f(|\gamma|)\equiv 1$.
Then
$$\sigma = \biggl[\int_0^1 d|\gamma|{2 N\,f(|\gamma|)\over |\gamma|^2+\Delta^2}
\biggr]^{-1/2}.\eqn\sigtwo$$
For the special case $f(|\gamma|)=1$ and $\Delta\ll 1$, equation \sigtwo\
can be evaluated in closed form: $\sigma = [\Delta/(\pi N)]^{1/2}$.  In general
we may write
$$\sigma = \biggl({\Delta\over \zeta\pi N}\biggr)^{1/2},\eqn\sigthree$$
where $\zeta$ is a correction factor.  Empirically, Mould et al.\ (1994)
find that the distribution of ellipticities is more skewed toward low
values than is a uniform distribution.  That is $\zeta>1$.  Using their
measured distribution, we estimate $\zeta\sim 2.2$.

	Finally, we suppose that the ellipticities of $N$ images have
been measured, each with accuracy $\Delta_j$.  Then
$$\sigma = \biggl(\zeta\pi\sum_{j=1}^N \Delta_j^{-1}\biggr)^{-1/2}
= \biggl({\overline{\Delta}\over 7 N}\biggr)^{1/2},\eqn\sigfour$$
where $\overline{\Delta}$ is the harmonic mean of
the $\Delta_j$ and where we have estimated $\zeta\pi=7$.

	Note that $\sigma$, the accuracy of the measurement of the components
of $\grk$, is proportional to the square root of $\Delta$, the accuracy of
the measurements of the ellipticities of the observed images.  This contrasts
sharply with the usual situation where the error in the determination of
a given parameter is directly
proportional to the errors in the measured quantities.

\chapter{Sensitivity of the Sloan Survey}

	The Sloan Survey  is a digital survey of $\pi$ steradians
about the north galactic pole.  The survey will be performed on a 2.5m
telescope in five bands by scanning the sky at approximately the sidereal
rate.  The nominal limit of the survey for a point source with
signal-to-noise ratio of 5 is
$r'=23.1$ (D.\ Weinberg 1994, private communication, DW).
Here we restrict consideration to galaxies with $r'<21.5$.  There are
$\sim 10^4$ such galaxies per square degree and these have a
median half-light diameter $\sim 2.\hskip-2pt ''5$ (DW).
It is difficult to assess
how well the ellipticities can be measured, but extrapolating from the
experience of Mould et al.\ 1994, we conservatively
estimate $\overline{\Delta}=0.07$.
We then estimate the sensitivity of the Sloan Survey from equation \sigfour\
to be
$$\sigma = 0.10\%\biggl({A\over {\rm deg}^2}\biggr)^{-1/2},\eqn\sigsloan$$
where $A$ is the angular area over which the mean value of $\grk$ is
being measured.

\chapter{Weak Lensing Signature of Nearby Structures}

	The Local Supercluster and other structures of the nearby universe
produce weak lensing effects $|\grk|\sim 0.2\%\Omega$ where $\Omega$ is
the density of the universe as a fraction of the critical density.  The
effects are coherent over tens of square degrees.
\FIG\one{Weak lensing by nearby structures in an $\Omega=1$ universe.
Galaxies are assumed to be isothermal spheres truncated at 2.8 Mpc, with
mass to blue light ratios of 2000.  The
lengths of the line segments in degrees are equal to magnitudes of the
mean polarization in units of 0.40\%.  The directions of the line segments
are the axes of
elongation.  The position (0,0) is the north galactic pole (NGP).  The Coma
cluster is $\sim 2^\circ$ from the NGP, the Virgo cluster is $\sim 15^\circ$
below it, and A1367 is $\sim 15^\circ$ below and to the right of the NGP.
The heart of the Local Supercluster runs from the Virgo Sourthern Extension
at roughly the lower-left corner through Virgo and out toward Ursa Major
which lies beyond the upper right corner.
}
\FIG\two{Same as Fig.\ \one\ except for an $\Omega=0.3$ universe with
the truncation radii of galaxies set to
0.84 Mpc, the mass to blue light ratio set to 600, and with $1^\circ$ line
segments representing 0.12\% polarization.}
This is apparent from Figures \one\ and \two\ which show the weak lensing
patterns for $\Omega=1$ and $\Omega=0.3$ universes respectively.  The
orientation of the line segments indicates the direction in which the images
are stretched.  The length of the segment indicates the size of the effect:
a 1 degree segment represents $|\xi|= 0.40\%$ in Figure \one\ and
$|\xi|=0.12\%$ in Figure \two.  To construct these figures, we computed the
deflection of light in many directions due to all the galaxies in
de Vaucouleurs et al.\ (1991 RC3) listed with redshifts and total
blue magnitudes.
We assumed that each galaxy has a total mass
to $B$-light ratio of $2000\Omega h$ where $h$ is the Hubble parameter
in units of $100\,\rm km\,s^{-1}\,Mpc^{-1}$ (Binney \& Tremaine 1987,
assuming $\VEV{B-V}_{\rm galaxies} \sim 0.8$),
and that the mass is
distributed in a truncated isothermal sphere with radius $2.8\Omega\,$Mpc.
We then found the polarization at $0.2^\circ$ intervals from the traceless part
(shear) of the
magnification tensor, and finally averaged the results over the 25 positions
within $1^\circ$ squares.

	The accuracy of the lensing pattern shown in Figures \one\ and \two\
depends on the assumption that the RC3 is complete, and also on the
correctness of the particular model we have chosen for the relation
between mass and light.  Both of these assumptions are likely to fail.
However, as we show below, the incompleteness of the catalog will be
rectified by the Sloan Survey itself, and the lensing pattern is mainly
sensitive to $\Omega$ rather than to the details of the correlation between
mass and light.

\section{Completeness}

	For the Sloan Survey, the median redshift of the galaxies with
redshifts will be $z=0.1$.  For galaxies with $r'<19.5$ the median
redshift will be $z=0.25$ (DW).  One can estimate the redshifts
for these latter from their colors.  The estimates should be fairly accurate
at least in a statistical sense.  Thus, it will be possible to make a
good estimate of the distribution of galaxies brighter than $L_*$ at least
out to $z=0.25$.  Under the assumption that the $L_*$ galaxies trace the
distribution of all galactic light, one can then predict the lensing due
to observed galaxies $z=0.25$ for a given model
relating mass to light.  That is, it will be possible to ``take out''
the effect of lensing due to galaxies at intermediate redshift, leaving
only the effects of the nearby structures and the galaxies with $z>0.25$.

	As we show explicitly in \S\ 2.3, below, the effect of lensing due
to galaxies with $z>0.25$ is expected to be small on the $\sim 5^\circ$
scales over which the effects of local structures are coherent.  Hence, these
distant galaxies can, to leading order, be ignored.

\section{Mass Models}

	Figures \one\ and \two\ look very similar to the eye.  The figures
would look exactly the same except that in Figure \one, we assumed that
the mass of galaxies is distributed to a radius of 2.8 $h^{-1}$Mpc, while
in Figure \two, we assumed 0.84 $h^{-1}$Mpc.  Figure \two\ represents a
universe with only 3/10 as much mass, but the line segments are 10/3 larger
for the same amount of lensing.  To the extent that Figures \one\ and \two\
look the same, one can measure $\Omega$ directly from the amplitude of the
observed lensing, without worrying about the details of the mass model.
To the extent that they are different, one can use the differences to
make inferences about the distribution of mass.  Comparison of the two
figures shows that it will be fairly easy to measure $\Omega$ using weak
lensing pattern averaged over many many square degrees.
On the other hand, information about the mass distribution will come primarily
from regions within a few degrees of clusters.

\section{Polarization Due to Distant Galaxies}

We can calculate the polarization field due to distant galaxies in a
given cosmological model, e.g.\ Cold Dark Matter (CDM).
To be specific, let us
assume that there is a sheet of galaxies at redshift $z_2$
corresponding to a comoving angular diameter distance
$x_2=2(1-(1+z_2)^{-1/2})$.
The derivations below follow Blandford et al.\ (1991) and Mould et al.\
(1994).  The derivations are formally
 valid only for $\Omega=1$ but can be easily rescaled to other values
of $\Omega$.  The polarization from all sources closer than $x_1$ is
$$
p(x_1,x_2)=2 \int_0^{x_1} dy {y(x_2-y) \over x_2}\calf_0(\vecy),\eqn\jone
$$
$$
\calf_0=\left({\partial^2 \over \partial y^2}-{\partial^2 \over \partial
x^2}-2i{\partial^2 \over \partial x \partial y} \right)\Phi_0,\eqn\jtwo
$$
where the line of sight is the z-axis
and $\Phi_0$ is the gravitational potential.
This real space formulation is suitable
for a known, or
assumed, mass distribution, but for a cosmological model it is more
useful to do a plane wave decomposition of the density field and then
specify the power spectrum of density fluctuations
$P(k)=|\delta_0^2(k)|$.
One can generate a map of the
polarization field as a random realization with the appropriate power
spectrum of polarization fluctuations $Q(k)$ which can be calculated
from $P(k)$.  Typically, the density field is assumed to be Gaussian
which means that the phases of the waves are uncorrelated and random.

The field for a sheet of galaxies at distance $x_2$ from sources closer
than $x_1$ is
$$\eqalign{&
p(\vecx_1,\vecx_2)=-3\int {d^3k \over {2\pi}^3}\int_0^{x_1} dy
y\left({x_2-y\over x_2}\right)\calf_0(\veck)\exp\left(i\ky\right)=\cr &
-{3\over 2}x_1^2\times\int {d^3k \over {2\pi}^3}\calf_0(\veck)
\left[{x_1 \over x_2}{j_1\left(\kx\right)\over \kx}+\left(1-{x_1\over
x_2}\right) \left(j_0\left(\kx\right)+i
j_1\left(\kx\right)\right)\right]e^{i\kx}.}\eqn\jthree
$$
where
$$
\calf_0(\veck)=\delta_0(\veck){(k_1+ik_2)^2\over k_1^2+k_2^2};\qquad
\kx\equiv {{\bf k\cdot x}_1\over 2}\eqn\jfour
$$
For $x_1 = x_2$ this reduces to the standard result from Blandford et al.\
(1991).  In the same way the polarization correlation
$\cpp(x_1,x_2,\theta)$ and the variance $\var(x_1,x_2,\theta)$ can be
calculated using the Fourier convolution theorem and the result
$$
\eqalign{&\int_{-\infty}^\infty da \exp(-iqa)\left[{x_1 \over x_2}
{j_1\left(a\right)\over a}+\left(1-{x_1\over x_2}\right)
\left(j_0\left(a\right)+ij_1\left(a\right)\right)\right]\cr = &
\pi\left[{x_1 \over x_2}{\left(1-q^2\right)\over 2}+
\left(1-{x_1 \over x_2}\right)(1-q)\right]\Theta(1+q)\Theta(1-q).}\eqn\jfive
$$
Then,
$$\eqalign{
\cpp&(x_1,x_2,\theta)=36\pi^2 x_1^3 \int_0^\infty dk k P(k)\cr &\int_0^1 ds
\left[\left({x_1\over x_2}\right)^2 s^2(1-s)^2+
\left(1-{x_1\over x_2}\right)s(1-s)\right]J_0(k x_1 \theta s),}\eqn\jsix
$$
and
$$\eqalign{
\var&(x_1,x_2,\theta)=36\pi^2 x_1^3 \int_0^\infty dk k P(k)\cr &\int_0^1 ds
\left[\left({x_1\over x_2}\right)^2 s^2(1-s)^2+
\left(1-{x_1\over x_2}\right)s(1-s)\right]
\left({2J_1(k x_1 \theta s) \over k x_1 \theta s}\right)^2.}\eqn\jseven
$$
The variance is interpreted as the mean square polarization when the
field is smoothed with a circular top hat weighting function of angle
$\theta$.  The power spectrum of polarization fluctuations is the
fourier transform of the polarization correlation function,
$$
Q(k,x_1,x_2)=18\pi x_1^3 \int_0^1 ds P\left({k \over
s}\right)\left[\left({x_1\over x_2}\right)^2(1-s)^2+\left(1-{x_1 \over
x_2}\right){1-s \over s}\right].\eqn\jeight
$$
A random realization of the polarization field from sources between
$x_1$ and $x_2$ can be generated by subtracting the field generated from
$Q(k,x_2,x_2)$ and $Q(k,x_1,x_2)$ using the same set of random numbers.
This is equivalent to specifying the same plane wave decomposition of
the density field $\delta_0(\veck)$ for the two polarization fields.

\FIG\three{Polarization field of galaxies at $z=0.40$ due to a CDM
$(h=0.5$,
$\Omega=1)$ mass distribution over the range $0.25<z<0.40$, smoothed over
1 square degree.  As in Fig.\ \one, a line segment of length $1^\circ$
represents a polarization of $0.40\%$.}
\FIG\four{Same as Fig.\ \three, except smoothed over 25 square degrees.
Direct comparison of this figure with Fig.\ \one, shows that the polarization
pattern induced by distant galaxies will not seriously interfere with
measurement of the field induced with local structures.}
The polarization field predicted for CDM generated by sources between
$z_1=0.25$ and $z_2=0.4$
is plotted in Figure \three\ smoothed on a scale of $0.56^\circ$.  This
corresponds to a smoothing area of 1 square degree.  It is plotted
in the same way as the predicted polarization field in Figure \one.  The
rms polarization of the polarization field generated by all sources is
2.5\% without smoothing and 1.1\% with this 1 square degree smoothing.
However, if we only look at sources more distant than $z_1=0.25$ then
the rms polarization drops to 0.34\% which is similar to what is
predicted from the local galaxies in the RC3 catalogue with $\Omega=1$.
This strong reduction of the
background signal comes about because the distant structures
generate mostly small angular scale structure in the polarization field.
Since the polarization field due to local structures is coherent over
many square degrees, the polarization induced by distant galaxies should
also be smoothed over a large area before comparing it with the locally
induced structure.  In Figure \four, we show the polarization due to distant
galaxies smoothed over 25 square degrees.  The rms polarization is $0.09\%$
on this scale and therefore should not seriously interfere with measurement
of the local structure.

\chapter{More Distant Structures}

	As we discussed in \S\ 1, it is particularly interesting
to measure the lensing due to local structures since we have the most
other information about them.  However, there is also much to be gained
by analyzing the lensing due to structures at intermediate redshifts
$z<0.25$.  First, of course, by simultaneously fitting the lensing amplitude
due to all the observed structures $z<0.25$ rather than just the local
structures, one could obtain a more accurate estimate of $\Omega$.  Second,
it is possible that a substantial part of the mass of the universe
is correlated with the light only on scales of many Mpc, or perhaps tens of
Mpc.  To probe these large physical scales effectively, it is necessary
to look at structures at larger distances.

\chapter{Systematic Effects}

	Since weak lensing by local structures is extremely weak
${\cal O}(0.2\%)$, one must be especially careful about small systematic
effects.  As discussed by Mould et al.\ (1994), these are basically of
two types.  First, trailing of the point
spread function (PSF) and second, classical aberration.
Mould et al.\ calibrated the trailing of
the PSF primarily by measuring the trend of the polarization with inverse
galaxy size, in effect extrapolating to galaxies of infinite size for
which a trailing PSF would have no effect.  They also checked for consistency
with the trailing measured from stellar images.  However, since there were
$\sim 4000$ galaxies and only $\sim 80$ stars, the galaxies provided a
somewhat more precise estimate than the stars.

	In the Sloan Survey, by contrast, there will be $N = 10^4$ galaxies
$\rm deg^{-2}$ compared to $\sim 2000$ stars $\rm deg^{-2}$.  This means that
the stars will provide much more information about the trailing of the PSF than
the galaxies.  It is easy to see that in fact the stars will provide adequate
information to calibrate the trailing.  The effective number of galaxies,
that is those that enter with significant statistical weight, is
$\zeta\pi\Delta N/2\sim 2500\,\rm deg^{-2}$.  The ellipticities of the stellar
images can be measured somewhat better than those of the galactic images,
but to be conservative we will assume equal accuracy.  Then, if there were
equal numbers of the stars and galaxies, the error in the galactic polarization
induced by the uncertainty in the measurement of
stellar trailing would be $\sigma$ times the
ratio of star-to-galaxy areas, i.e., $\sim 16\%\sigma$ for a $1''$ seeing
disk and median $2.\hskip-2pt'' 5$ diameter galaxies.  Since there
are slightly more effective galaxies than stars, this fraction is raised
by $\sqrt{1.25}$ to $\sim 18\%\sigma$.
In other words, the problem of calibrating the trailing of the PSF increases
the error in the polarization measurement by a small fraction.

	Mould et al.\ (1994)
calibrated the classical aberration using photometry
of astrometric fields.  We assume that calibration for the Sloan Survey
can be carried out in a similar manner.
It is possible that flexure of the telescope will
alter the classical aberration as a function of the orientation of the
telescope.  If so, this would complicate the calibration.

	Finally, we note that the polarization measurements themselves will
provide an important check on how well the systematic effects have been
calibrated.  The polarization will have significant power on scales of
tens of square degrees due to local structures, but not on scales of
steradians.  Hence, even a very small systematic distortion induced by the
instrumental setup should be recognizable.  Moreover, since the pattern
of weak lensing due to local structures is approximately known and
relatively complicated, it should be possible to track down systematic
effects from unanticipated sources.  That is, systematic effects from
any source would be unlikely to mimic this general pattern.

{\bf Acknowledgements:  }We would like to thank David Weinberg for providing
us with a summary of the expected characteristics of the Sloan Survey sample,
and Robert Blum for making several helpful suggestions.

\endpage

\Ref\aa{Binney, J.\ \& Tremaine, S.\ 1987, Galactic Dynamics, p.\ 617
(Princeton: Princeton Univ.\ Press)}
\Ref\aa{Blandford, R.\ D., Saust, A.\ B., Brainerd, T.\ G.,
\& Villumsen, J.\ V.\ 1991, MNRAS, 251, 600}
\Ref\aa{de Vaucouleurs, G., de Vaucouleurs, A., Corwin, H.\ G.\ Jr.,
 Buta, R.\ J., Paturel, G.\ \& Fouqu\'e, P.\ 1991, Third Reference Catalogue
of Bright Galaxies, (New York: Springer-Verlag)}
\Ref\aa{Fort, B., Le Borgne, J.\ F., Mathez, G., Mellier, Y., \& Picat, J.\
P.\ 1991, Annales de Physique, 16, 211}
\Ref\aa{Gunn, J.\ E.\ 1967, ApJ, 150, 737}
\Ref\aa{Gunn, J.\ E.\ \& Knapp, G.\ R.\ 1993, Astronomical Surveys, B.\ T.
Stoifer, ed., ASP Conference Proceedings 43, 267}
\Ref\aa{Kaiser, N.\ 1992, ApJ, 388, 272}
\Ref\aa{Kristian, J.\ 1967, ApJ, 147, 864}
\Ref\aa{Lynds, R.\ \& Petrosian, V.\ 1989, ApJ, 336, 1}
\Ref\aa{Miralda-Escud\'e, J.\ 1991, ApJ, 380, 1}
\Ref\aa{Mould, J., Blandford, R., Villumsen, J., Brainerd, T.,
Smail, I., Small, T., \& Kells, W.\ 1994, MNRAS, submitted}
\Ref\aa{Soucail, G., Fort, B., Mellier, Y., \& Picat, J.\ P.\ 1987, A\&A, 172,
L14}
\Ref\aa{Tyson, J.\ A., Valdes, F., \& Wenk, R.\ A.\ 1990, ApJ, 349, L1}
\Ref\aa{Valdes, F., Tyson, J.\ A., \& Jarvis, J.\ F.\ 1983, ApJ, 271, 431}
\refout
\endpage
\figout
\endpage
\bye
\magnification=\magstephalf

\bye